\documentclass{aa}
\usepackage{sidecap}  
\usepackage{graphicx}
\usepackage{txfonts}
\usepackage[colorlinks = true,
            linkcolor = blue,
            urlcolor  = blue,
            citecolor = blue]{hyperref}

\begin{document} 

\newcommand{\myrev}[1]{#1}
\newcommand{\secondrev}[1]{#1}

   \title{Forming equal-mass planetary binaries via pebble accretion}

   %\subtitle{I. Subtitle here}

   \author{T.J. Konijn$^1$\thanks{Corresponding author: \href{mailto:tom@nsaweb.nl}{tom@nsaweb.nl}},
          %\inst{1}
          R.G. Visser$^1$,
          C. Dominik$^1$,
          C.W. Ormel$^2$
          }

   \institute{$^1$ Anton Pannekoek institute for Astronomy (API), University of Amsterdam, Science Park 904, 1098XH Amsterdam \\
   $^2$ Department
of Astronomy, Tsinghua University, 30 Shuangqing Rd, 100084 Beĳing, PR China}

   \date{Received 10 October 2022 / Accepted 29 November 2022}

% \abstract{}{}{}{}{} 
% 5 {} token are mandatory
 
  \abstract
  % context heading (optional)
  % {} leave it empty if necessary  
  {Binary Solar System objects are common, ranging from satellite systems with very large mass ratios, $M_1/M_2$, to those with mass ratios approaching unity. One well-known example of a binary is the Pluto-Charon system. With Charon being 'only' eight times less massive than Pluto, the question arises (as in the case of many other systems) as to why the mass ratio is still close to unity. There is much evidence that (binary) planet(esimal) formation happened early, when the protoplanetary gas disk was still present. It is likely that (at least some of) these binaries  evolved together, as a result of pebble accretion. 
  Pebble accretion is a  new key paradigm in planetary formation and it is believed to play a major role in many aspects of the formation of planetary systems, from the radial transport of material to the rapid growth of planetary embryos throughout the system.}  % aims heading (mandatory)
   {Here, we focus on the question of how the mass arriving in the gravitational influence zone of the binary during pebble accretion is distributed over the binary components for a given initial mass ratio. We also consider whether accretion over time leads to equal-mass binaries (converging mass ratio) or to a dominant primary mass with a small moon (diverging mass ratio).}
  % methods heading (mandatory)
   {We numerically integrated two-dimensional (2D) pebble trajectories in the same typical fashion as for a single mass that is subject to pebble accretion. We tracked the efficiency of accretion for the two separate binary components, compared to a single body with the same mass. These numerical simulations were done for a range of binary mass ratios, mutual separations, Stokes numbers, and two orbital distances, 2.5 and 39 au.}  % results heading (mandatory)
   {We find that in the limit where pebbles start to spiral around the primary (this holds for relatively large pebbles), the pebble preferentially collides with the secondary, causing the mass ratio to converge towards unity. In our tested case, where the total binary mass is equal to that of the Pluto-Charon system, this takes place on $\sim$Myr timescales. In this regime the total sweep-up efficiency can lower to half that of a pebble-accreting single body because pebbles that are thrown out of the system, after close encounters with the system. These timescales and sweep-up efficiency are calculated under the assumption our 2D simulations compare with the 3D reality. The results show that systems such as Pluto-Charon and other larger equal mass binaries may well have co-accreted by means of pebble accretion in the disk phase without producing binaries, with highly diverging mass ratios.}
  % conclusions heading (optional), leave it empty if necessary 
   {}

   \keywords{Pebble Accretion -- 
            Binary Planetesimals -- 
            Planetary formation -- 
            Streaming instability
               }

   \maketitle
%
%-------------------------------------------------------------------

\section{Introduction}
Roughly twenty percent of the cold classicals in the Kuiper belt are binary systems \citep{2008Icar..194..758N}, with the mass ratio of the components being close to unity. A well-known example is the Pluto-Charon system \citep{1978AJ.....83.1005C} located at an orbital distance from the Sun of around 39 au. With Pluto being roughly eight times as massive as Charon, it is often categorised as a nearly equal-mass binary. Recent studies have shown,  with increasing certainty, that equal-mass binaries are not rare in the Kuiper Belt and their occurrence rate is estimated to be of at least several percent \citep{2008Icar..194..758N} \footnote{The authors classify it as "equal size binaries". The internal densities are very similar for cold classicals making it equivalent to the definition of equal mass binaries.}. 

In addition, the asteroid belt features numerous objects with companions \citep{2002aste.book..289M}.  Some of these binaries have similar masses (e.g. 90 Antiope, 2006 VW139, 2017 YE5, and 69230 Hermes \citep{2008Icar..196...97M,2004DPS....36.4602M,Agarwalletal2017,2012Icar..221.1130M}), while others have large mass ratios, so that calling the smaller objects asteroid moons or satellites and the larger ones planet or primary would seem more correct. 

The Earth-Moon system is a binary with a relatively large mass ratio as well, although here the definition of equal mass binary reveals its ambiguity. The most recent definition of 'equal-mass' and 'near-equal-mass' reveals the need for the barycenter to lie \secondrev{outside} both companions \citep{AlanLevison2002}. Another way of categorising the mass equality of binaries is the 'tug-of-war' ratio, expressing the planetary gravity over the solar gravity for a satellite. Equal-mass planet binaries are often also called 'double planets', making the definition of an equal-mass binary even more vague; the Moon ticks off all the criteria for being a planet on its own and the tug-of-war ratio for the Earth-Moon system is $0.46$ \citep{Herschel}, but the barycenter lies within the body of the Earth making the Earth-Moon system: a double-planet system, an equal-mass binary, or a primary (Earth) and secondary (Moon) system, respectively \citep{Russel2017}. 

The formation of binary planets is still an open question and a number of channels have been proposed. In the classical model it is thought, at least in the case of Pluto-Charon \citep{2005Sci...307..546C,2011AJ....141...35C} and Earth-Moon systems, that they formed via giant impacts on the primaries, smashing debris into stable orbiting secondaries \citep{1976LPI.....7..120C,2012Sci...338.1052C}. This scenario requires specific impact parameters that ensure sufficient angular momentum in the ejected debris  to allow for the formation of an object outside the Roche limit.

A second viable mechanism to create binary planet(esimals) is capture via three-body interactions in an early, dense phase of the solar system, where such interactions were much more frequent than they are today \citep{Goldreichetal2002,2005MNRAS.360..401A,2007MNRAS.379..229L,2008ApJ...673.1218S}. This might be the case for the dynamically hot classicals in the Kuiper Belt, which are more scattered due to perturbations by Neptune and have relatively wide mutual separations. They also show a broad and more chaotic distribution in orbital parameters, are subject to Kozai-Lidov effects \citep{Kozai,Lidov1962}, and show a much lower equal mass binary occurrence than the cold classical population because disruption is relatively easy \citep{Morbinesv2020}. The objects in this population thus have a dynamical profile that fits a random capture process which suggests a different formation process than the cold classicals. The cold classicals have strong color correlations, indicating the same composition \citep{Benechietal2009}, they have mass ratios close to unity, and show a clear preference for prograde orbits (80 \%) versus retrograde orbits (20 \%) \citep{Grundyetal2019}, strongly disfavouring a capture event.

Finally, a recently proposed mechanism to explain the formaton of planetary binaries is the streaming instability (SI) \citep{Youdin_2005,2011EM&P..108...39J,2021A&A...647A.126V}. In this scenario, dust clumps together via an interaction with the disk gas and when it becomes gravitationally bound, it collapses on a dynamical timescale to form comets, planetesimals, and protoplanets \citep{2010AJ....140..785N,2019NatAs...3..808N}. Depending on the amount of angular momentum present in the clump before the collapse \citep{2021PSJ.....2...27N}, and possibly added to it intrinsically during the collapse \citep{VisserandBrouwers2022}, the forming object will become a binary in order to absorb excess angular momentum. In this scenario, the disk is expected to be loaded with pebble-sized objects, with Stokes numbers between $10^{-2}$ to unity. 

Regardless of the process of formation, if these binary systems formed early in the disk, their further evolution and growth is non-trivial. The presence of nebular gas and/or solids might cause the binary components to spiral closer and merge if they are aerodynamically small within a Myr \citep{Lyraeal2021} or start wide and merge through secular effects \citep{Rozneretal2020}. However, if the binary components form early and start massive enough  for gravity to dominate, they might continue to grow through co-accretion. The growth of such a binary system is widely known to lead to a convergence in the mass ratio -- if the satellite and primary accrete from a planetesimal swarm under the right conditions. A satellite would grow significantly faster than its primary due to its higher surface-to-mass ratio and by exploiting the gravitational focusing of the primary \citep{HarrisKaula1975}, provided that the mutual separation is small enough \citep{MorWat2001}. 

While this co-accretion scenario for binaries has been studied for planetesimal accretion, it has not been applied in the case of pebble accretion. In pebble accretion, massive bodies efficiently accrete solids through the interplay of gravity and the presence of nebular gas \citep{Ormel_Klahr_2010,LJ2012}. As the gas drag removes relative velocity and angular momentum from the solids, they are efficiently captured into the gravitational potential, leading to largely enhanced accretion cross-sections. Pebble accretion has only been invoked explicitly for a single pebble-accreting body to explain many observational difficulties in planet formation with great success, such as the size distributions of asteroids and Kuiper Belt objects  \citep{Johansenetal2015}, formation processes in the cores of gas giants \citep{Levisonetal}, TRAPPIST system architecture \citep{ormel-trappist,Schormel2019}, spin axis of solar system bodies \citep{Visser_2020}, formation of terrestrial planets \citep{Johetal2021}, and many other applications.

\myrev{Of course, pebble accretion will not be relevant for all planetary binary objects. The Earth-Moon system formed later, after the disk gas had already been dispersed -- so pebbles would have no longer been drifting inwards and the accretion of small rocks would not be enhanced by the interaction with the gas. Also, low-mass binary objects, as in the case of most Kuiper Belt binaries, will not have sufficient gravity to get pebble accretion started in a serious way. The process is, however, applicable for bodies that are sufficiently massive and form so early that the gas disk and pebbles are still around.}

In this study, we investigate the evolution and growth of a planetary binary subject to pebble accretion. We subdivide the total mass, $M,$ of a single body in two masses, $M_1$ and $M_2$, such that $M = M_1 + M_2$ with an increasingly diverging mass fraction, $f_m = M_1 / M_2 \in [1,8]$. We determine the timescale to $e-$fold the mass of the binary compared to the growth time of a single body with the same mass. We then quantify the $e$-folding time needed to converge the mass ratio to unity by looking at the individual efficiency of accretion from the flux of pebbles swept up by the system. In many cases we find that, in the 'die-hard' pebble accretion regime, the lower mass body (in this case $M_2$) grows faster than the more massive one in essence for the same reasons as in the planetesimal accretion case described above. As a result, the mass ratio can converge back to unity, well within the disk lifetime of $\sim$1 Myr. Pebble accretion provides an explanation for the significant fraction of (near) equal mass observed planetary binaries. 

The paper is organised as follows . In Section \ref{sec:1}, we discuss the model setup and the assumptions we use.  In Section \ref{sec:2}, we describe the results of the numerical simulations. In Section \ref{sec:3}, we discuss the outcomes and results, after which we summarise the implications and present our conclusions in Section \ref{sec:4}.
%--------------------------------------------------------------------
\section{Setup and assumptions}
\label{sec:1}
\subsection{Disk model}

In order to see how these binary systems evolve, we begin with a general description of the planet-forming disk in which the binary resides. For the gas temperature and surface density, we assume power-law profiles given by \citep{WeidenschillingB} and \citep{Hayashi1985}:
\begin{align}
        \label{eq:T} T(r_0) &= 170\ \mathrm{K}\left(\frac{r_0}{1\ \mathrm{ au}}\right)^{-1/2},\\
        \label{eq:Sigma}\Sigma(r_0) &= 1700\ \mathrm{g\ cm}^{-2} \left(\frac{r_0}{1\ \mathrm{au}}\right)^{-3/2},
\end{align}
where $r_0$ is the radial distance to the central star. We assume hydrostatic equilibrium, whereby the vertical density structure forms as a Gaussian:
\begin{align}
    \label{eq:rho} \rho(r_0, z) &= \frac{\Sigma(r_0)}{\sqrt{2\pi}H(r_0)}e^{-\frac{1}{2}\left(\frac{z}{H}\right)^2},
\end{align}
with $H = c_\mathrm{s}/\Omega_0$ as the scale height, $\Omega_0$ is the Keplerian frequency, $c_\mathrm{s} = \sqrt{k_\mathrm{B}T/\Bar{m}}$ is the local isothermal sound speed, $k_\mathrm{B}$ as the Boltzmann's constant, and $\Bar{m}$ the mean molecular weight. The gas moves at a slightly lower velocity than Keplerian, $v_\mathrm{K} = \sqrt{GM_\star/r_0}$ where $G$ is the gravitational constant and $M_\star$ is the mass of the star. \citep{WeidenschillingA} gives a dimensionless constant relating $v_\mathrm{K}$ and the headwind particles travelling in Keplerian orbit feel. This dimensionless constant, $\eta,$ can be estimated by looking at the $r$-dependencies of Eqs. \ref{eq:T}-\ref{eq:rho}:
\begin{align}
    v_\mathrm{hw} &= \eta v_\mathrm{K},\\
    \eta &= \frac{r_0}{2v_\mathrm{K}^2\rho}\frac{\mathrm{d}P}{\mathrm{d}r_0} = -\frac{13c_\mathrm{s}^2}{8v_\mathrm{K}^2},
\end{align}
where $v_\mathrm{hw}$ is the headwind and $P$ the pressure governed by the ideal gas law. For the adopted disk profile, the headwind has a numerical value of $v_\mathrm{hw} = -32$ m s$^{-1}$ downward in the co-moving local frame. In our parameter study, the Epstein regime for pebble stopping time is the relevant one and we calculated it from the Stokes number ($\tau_\mathrm{s}$), relating the stopping time ($t_\mathrm{s}$) to one orbital timescale:
\begin{equation}
    \tau_\mathrm{s} = t_\mathrm{s} \Omega_0.
    \label{eq:stok}
\end{equation}

\begin{table*}
\centering
\label{Tab:tablepars}
\caption{Overview of the parameter study used in the simulations. The ranges between brackets are logarithmically spaced containing 30 values for $\tau_\mathrm{s}$ and 10 values for $a_\mathrm{b}$. Since we are simulating two different mass ratios, $f_m$, at two different orbital distances $r_0$, we have 30x10x2x2 = 1200 simulations in total. The boldfaced quantities indicate the parameter set of the fiducial model.}
\tabcolsep=0.1cm
\begin{tabular}{llllll}
\hline
\hline
$r_0$ {[}au{]} & $\tau_\mathrm{s}$                                & $a_\mathrm{b}$ $[R_\mathrm{H}]$              & $f_m$ & N    & $M_\mathrm{total}$                     \\ \hline
2.5            & {[}$10^{-2}$-0.4{]}                              & {[}0.0075-0.05{]}                            & $2,8$    & 8100 & $M_\mathrm{Pluto} + M_\mathrm{Charon}$ \\
$\mathbf{39}$  & {[}$5 \times 10^{-4}$-0.1{]}, $\mathbf{10^{-2}}$ & {[}$10^{-3}$-$10^{-2}${]}, $\mathbf{0.0025}$ & $2,\mathbf{8}$    & 8100 & $M_\mathrm{Pluto} + M_\mathrm{Charon}$ \\ \hline

\end{tabular}
\end{table*}

\subsection{Pebble and binary dynamics}

\begin{figure}[!tp]
    \centering
    \includegraphics[width=0.45\textwidth]{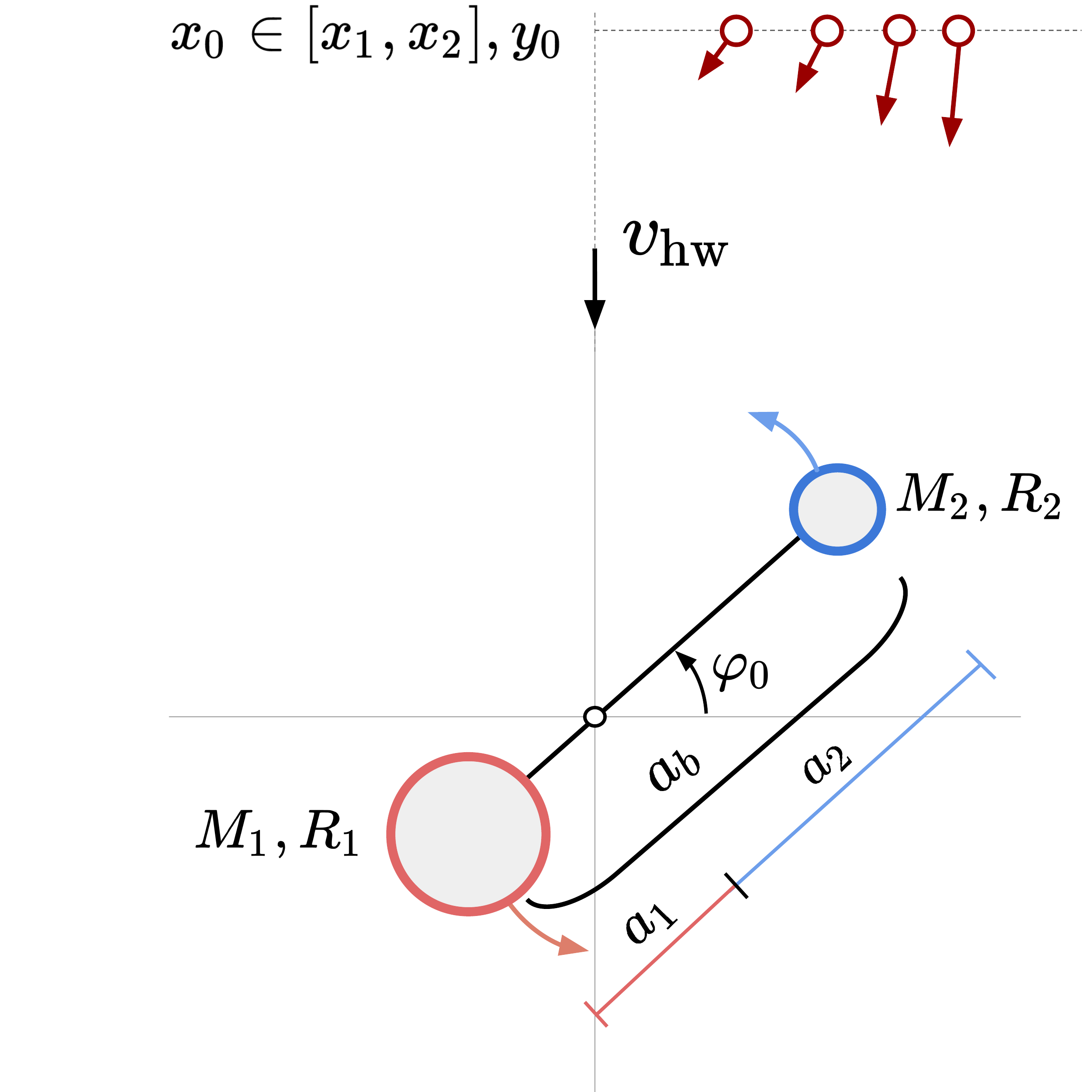} 
    \caption{Overview of the binary planetesimal system in the co-moving frame. The star is located on the left side of the frame. The binary components are denoted with their masses, $M_1,M_2$, and radii, $R_1, R_2$, travelling around their barycenter in circular orbit at distance, $a_1$ and $a_2$, respectively. Pebbles are released from a distance $y_0$ between the edges of the impact range, $x_1,x_2$, for a given initial phase angle of the binary, $\varphi_0$, with respect to the $x$-axis. The starting velocities of different pebbles differ depending on the $x$-coordinate as seen in Equation \ref{v_y}, therefore, their velocity vectors point in different directions. The background gas headwind is denoted with the black arrow on the y-axis.}
    \label{fig:overview}
\end{figure}

We adopted a two-dimensional (2D) local shearing sheet box co-moving with the binary center-of-mass (barycenter) at distance, $r_0$, from the central star. In this frame the equation of motion of a pebble is given by \citep{Ormel_Klahr_2010}:
\begin{align}
    \frac{\mathrm{d} \mathbf{v}}{\mathrm{d} t} = \begin{pmatrix}
    2 \Omega_0 v_y + 3\Omega_0^2x\\ 
    -2\Omega_0 v_x
    \end{pmatrix} -\sum_{j=1}^{2} \frac{G M_{j}}{|\mathbf{r} - \mathbf{r}_j(t)|^3}\begin{pmatrix}
    x - x_j(t)\\
    y-y_j(t)
    \end{pmatrix} + \mathbf{a}_\mathrm{drag},
\end{align}
with the first expression in brackets consisting of the Coriolis and tidal accelerations, the second term with the two-body gravity of the separate binary components with mass, $M_j$, and Cartesian positions ($x_j(t)$, $y_j(t)$). 

From here on we use the following terminology for the binary components. We say the binary is equal mass if the mass ratio is unity and no explicit reference to one of the bodies has to be made due to symmetry. If the mass ratio changes to $f_m \equiv M_1/M_2 > 1$ we say that the more massive one with mass, $M_1$, is the 'primary', and the least massive one with mass, $M_2$, is the 'secondary'.

We consider a binary separation that is narrow and we assume mutual circular motion of the binary system during the accretion. The circular motion of the system with total mass, $M = M_1 + M_2$, is then modelled over time, $t,$ with:
\begin{equation}
    \begin{pmatrix}
    x_{1,2}\\ 
    y_{1,2}
    \end{pmatrix}= a_{1,2}\begin{pmatrix}  \cos (\varphi_0 + \omega t)\\\sin (\varphi_0 + \omega t) \end{pmatrix},
    \label{eq:phase}
\end{equation} 
where $\omega = \sqrt{G M / a_\mathrm{b}^3}$ as the angular frequency, $\varphi_0$ as the initial phase angle with which the simulation started and the subscripts indicating to which body it applies, respectively. The binary components are assumed massive and unaffected by the gas. They are placed on a circular orbit around the barycenter with orbital separation $a_\mathrm{b} = a_1 + a_2$ and, if $f_m \ne 1$,  with $a_1$ as the distance of the primary with mass, $M_1$, and radius, $R_1$, and $a_2$ as the distance of the secondary with mass, $M_2$, and radius, $R_2$, to the barycenter. The system rotates in a prograde orientation around its barycenter. 

The Cartesian coordinates and velocities of the pebble are given by ($x$, $y$, $v_x$, $v_y$). The last term contains the acceleration due to the drag force on the pebble equal to:
\begin{equation}
    \mathbf{a}_\mathrm{drag} = 
    -\frac{\mathbf{v} - \mathbf{v}_\mathrm{g}}{t_\mathrm{s}},
    \label{eq:drag}
\end{equation} 
with $\mathbf{v}_\mathrm{g}$ the gas velocity including shear given by:
\begin{equation}
    \mathbf{v}_\mathrm{g} =  \left(v_\mathrm{hw} - \frac{3}{2}\Omega_0 x\right) \mathbf{e}_y.
    \label{eq:vgshear}
\end{equation}
An illustrative sketch of the setup is shown in Figure \ref{fig:overview}.  

The total mass, $M,$ of the system in our model corresponds to the mass of the Pluto-Charon system, with an internal density $\rho_{\bullet} = 1.778\,\mathrm{g \ cm^{-3}}$ taken to be equal for both bodies. The (shared) Hill radius is then given by:
\begin{equation}
    R_\mathrm{H} = r_0 \left(\frac{M}{3 M_\mathrm{\star}}\right)^{1/3}.
    \label{eq:Rh}
\end{equation}
In varying the mass ratio of the binary system, we then subdivided the total mass accordingly over the components from which their barycentric distance, $a_1$ and $a_2$, directly follow as:
\begin{equation}
    a_{1,2} =\frac{M_{2,1}}{M} a_\mathrm{b}.
    \label{eq:barya}
\end{equation}
From the assumed internal density the radius of each body is then calculated from:
\begin{equation}
    R_{1,2} = \left(\frac{3M_{1,2}}{4 \pi \rho_{\bullet}}\right)^{1/3}.\label{eq:radbodies}
\end{equation}

\subsection{Numerical method and initial conditions}
\label{subsec:nummethod}
We are interested in the (relative) growth rate of the binary (components) subject to pebble accretion. To do so we integrate the pebble trajectories with the Runge-Kutta-Fehlberg step variable integration method \citep{Fehlberg69,RKF45}, with a relative error tolerance of $\mathrm{tol} = 10^{-6}$. The pebbles are released on a distance far away from the binary in the y-direction:
\begin{align}
        y_0 = C\sqrt{\frac{GMt_\mathrm{s}}{v_\mathrm{hw}}},
        \label{eq:y0}
\end{align}
(as illustrated in figure \ref{fig:overview}) with $C = 200$, a safety factor. This ensures that, initially, gas drag is dominant over the gravity from the binary center-of-mass. The initial velocities of the incoming pebbles are, hence, given by the unperturbed radial and azimuthal drift velocities \citep{WeidenschillingA}:
\begin{align}
        \label{v_x} v_{x,\infty} &=- \frac{2\tau_\mathrm{s}}{\tau_\mathrm{s}^2 + 1} v_\mathrm{hw},\\
        \label{v_y} v_{y,\infty} &= -\frac{1}{\tau_\mathrm{s}^2+1}v_\mathrm{hw} - \frac{3}{2}\Omega_0x_0,
\end{align}
where $-\frac{3}{2}\Omega_0x_0$ is the correction to the Keplerian Shear. The pebbles are released with a shear corrected sampling function, equivalent to the method used in \citet{Visser_2020}. In the co-moving local frame we adopt, the pebble flux released then increases for increasing $x_0$, to account for the increasing relative velocities of the pebbles with respect to the binary barycenter due to shear.

We then use a bisection algorithm to find the range of initial x-coordinates for which the pebbles hit one of the binary components, this range is verified to be equal to the impact parameter of a single planetesimal with mass, $M=M_1+M_2$, in the parameter space we investigate. We do note that this only is valid for the close binary separations we consider. A pebble is registered as a hit when it satisfies the condition:
\begin{align}
        \sqrt{(x - x_{1,2})^2 + (y - y_{1,2})^2} \le R_{1,2},
        \label{eq:hit}
\end{align} 
with the subscript again referring to the corresponding body. After the cross-section edges have been found, we release pebbles between the edges $x_0 \in [x_1,x_2]$ accounting for an error of missing trajectories of $\sim$ 5 percent. The resolution in pebbles released is taken to be $N_{x_0} = 90$ for each initial phase angle linearly spaced between $\varphi_0 \in [0,2 \pi)$ of the binary system with resolution $N_{\varphi_0} = 90$. In total, we then released the amount of $N_{x_0} \times N_{\varphi_0} = 8100$ pebbles per individual simulation. Finally, in each simulation, we summed the total number of accreted pebbles on each body $N_{1,2}$ over each initial phase angle and we defined the accretion ratio to be:
\begin{equation}
        \varepsilon \equiv \frac{\dot{M_1}}{\dot{M_2}} = \frac{N_1}{N_2},
        \label{eq:accratio}
\end{equation}
with $N_1$ and $N_2$ as the total hits on the primary or secondary, respectively, but when the mass ratio is not unity.

To structure the results from our parameter study, we defined a fiducial model based on the Pluto-Charon mass(-ratio), radii, orbital distance from the star $r_0 = 39$ au and separation distance $a_\mathrm{b} = 0.0025R_\mathrm{H}$, with a pebble Stokes number of $\tau_\mathrm{s} = 10^{-2}$. An overview of the fiducial model and the rest of our parameter study is provided in Table \ref{Tab:tablepars}. A reproduction package of the code can be found online\footnote{The reproduction package can be found on \href{https://doi.org/10.5281/zenodo.7324045}{doi.org/10.5281/zenodo.7324045}}.
    
\begin{figure*}[tp!]
    \centering
    \includegraphics[width=0.9\textwidth]{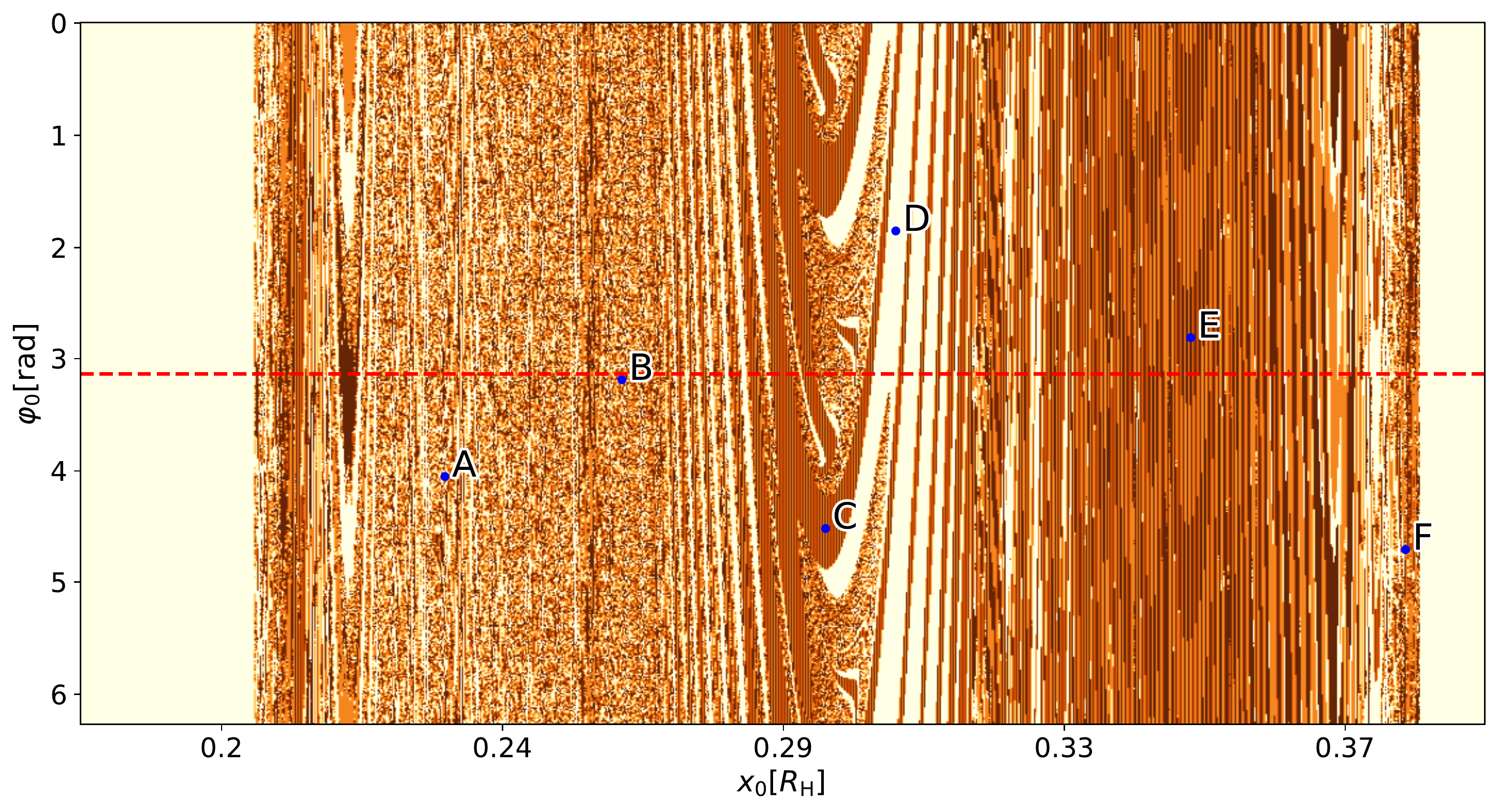}
    \includegraphics[width=1.0\textwidth]{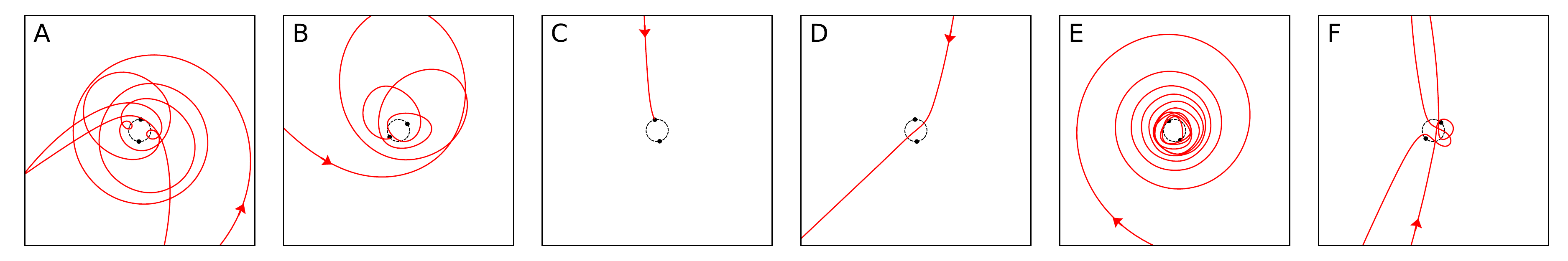}
    \caption{Hitmap of pebble impacts on an equal mass binary system such that $M_1=M_2=(M_\mathrm{Pluto}+M_\mathrm{Charon})/2,   $ shown at the top. Every pixel is a pebble falling into the system. On the $x$-axis the $x_0$ coordinate at moment of release is shown in Hill radii, while on the $y$-axis the initial phase orientation of the binary $\varphi_0$ in radians is depicted as illustrated in figure \ref{fig:overview}. The orange pixels show impacts on $M_1$, the brown pixels on $M_2$ \& the beige pixels indicate ejected pebbles. The final accretion ratio $\varepsilon = 1$, meaning that both bodies accrete an equal flux of pebbles because of symmetry arguments. The dashed red line indicates a starting angle of $\varphi_0=\pi$ rad. From 6 results in this graph the trajectories have been plotted below, those are denoted as A-F in this hitmap. The resolution has been taken $(N_{\varphi_0} \times N_{X_0} = 500 \times 1000)$. Bottom panel: Six trajectories from the hitmap above. Shown is a top-down view of the binary system with a pebble "falling" in. In each of the trajectories, an arrow denotes the pebble coming into the frame. Pebble trajectories A through F are sorted with their starting distance, $x_0$, going from left to right, corresponding to closer radial distance from the star to further away, respectively. We note that even though the trajectory for F appears to start on the bottom left, it originated on the right outside of the frame only entering the image below.}
    \label{fig:hmf}
\end{figure*}

\begin{figure*}[t]
    \centering
    \includegraphics[width=\textwidth]{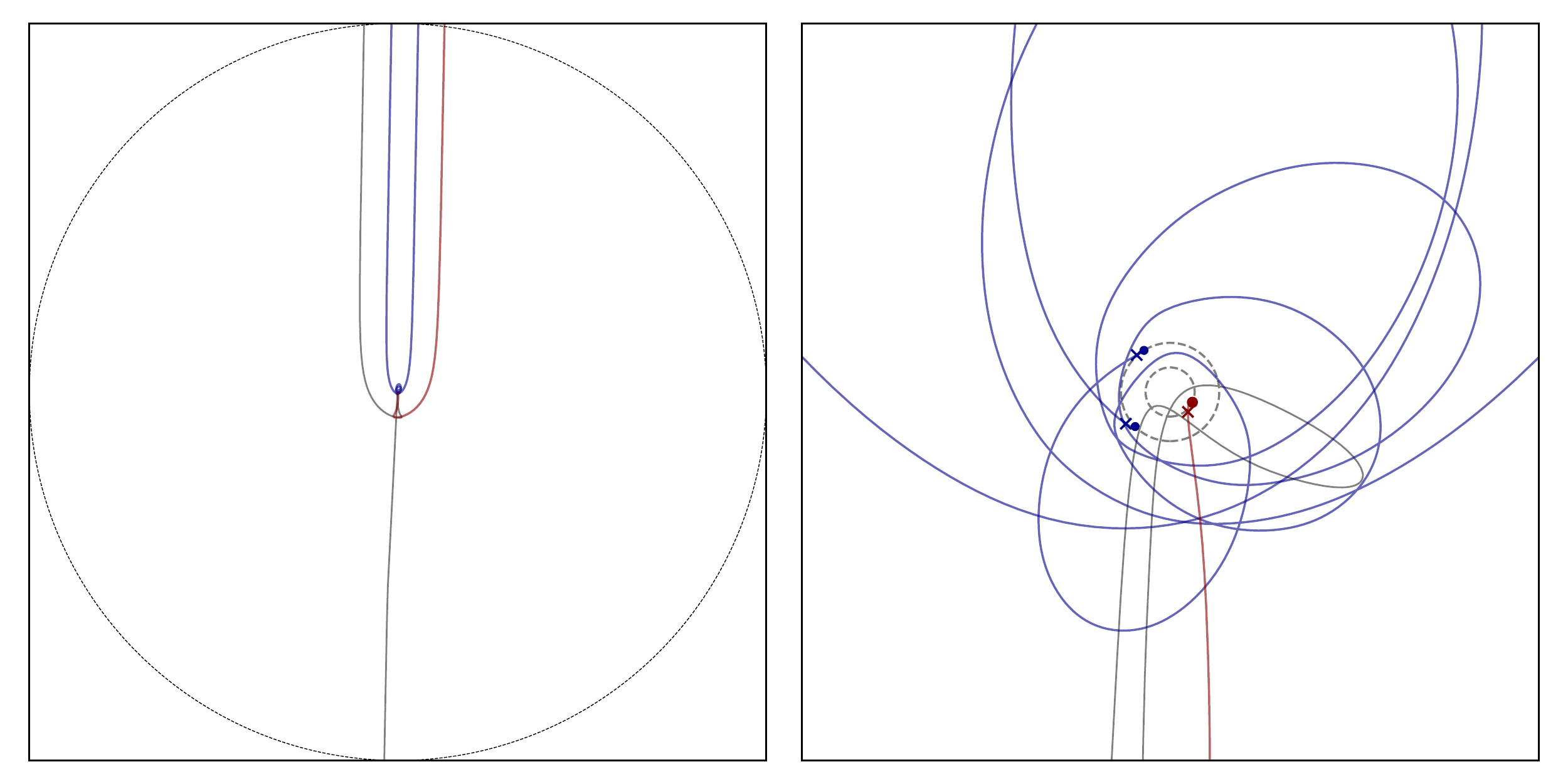} 
    \caption{Selection of pebble trajectories for a given initial phase angle for the fiducial model with a mass ratio of 2, to illustrate the accretion advantage for the secondary. Left panel: Pebbles enter the Hill radius (dashed circle) from the top and approach the binary center of mass in the middle of the panel. Gray trajectories miss the entire system, blue ones accrete onto the secondary and red ones onto the primary. Right panel: Zoom in on the trajectories shown in the left panel. The hits on the secondary are indicated with the blue crosses at point of impact. It clearly illustrates the accretion advantage for the smaller body due to close encounters and spirals. Additionally, we show the more massive body in the center (dark red solid circle) with a direct hit of a pebble (dark red trajectory). The trajectories are from different times and are plotted in one panel only for illustration.}
    \label{fig:fidacc}
\end{figure*}

\section{Results}
\label{sec:2}
\subsection{General interpretation}
We describe above the way in which a single pebble interacts with a binary planetesimal system, however, that does not explain how that system of planetesimals is expected to grow over time. For that purpose, we need every starting $x_0$-position where released pebbles will interact with the system and all starting angles $\varphi_0 \in [0,2 \pi),$ as described in Section \ref{subsec:nummethod}. In Figure \ref{fig:hmf}, we plot a hitmap of 500.000 pebbles using our fiducial system from Table \ref{Tab:tablepars}, with the only exception the mass ratio is set to $f_m=1$. Because of symmetry considerations, the accretion ratio $\varepsilon$ (Equation \ref{eq:accratio}) in this run is therefore unity.

Every pixel in the hitmap is a single pebble falling into the binary system, set by the $x$-coordinate when released, and the phase of the binary at that moment. When a pebble falls toward the system, there are three possible outcomes; the pebble can fall onto either of the planetesimals or it can fly by without hitting any of the two bodies and continue on its path around the central star. In the first two cases, the pixel is coloured either brown or orange depending on onto which body it accretes.  If the trajectory does not lead to accretion, it is coloured in beige.  This specific case describes an equal-mass binary, $M_1=M_2$, the planetesimals are moving as a mirror image of each other around the barycenter. Therefore, we expect that the top and the bottom half of the hitmap are identical, with inverted colours. This is indeed the case in Figure \ref{fig:hmf}. A horizontal dashed red line is added at $\varphi_0 = \pi$ to emphasise this symmetry. The exact same image can be seen above and below this line, only with the brown and orange colours swapped. In a system of a non-equal mass binary, this symmetry is expected to disappear.

By looking at the different trajectories found in the hitmap shown above, we can identify (at the simplest) three regions for all pebbles falling into a binary system, ranging from the most outer left to the most outer right. These three regions are as follows. 

In the first, most pebbles falling in from the left side of the map enter the system circling in a prograde direction around the binary, in the same direction as the planetesimals are moving. This can result in a pebble spiralling into the system slowly (as in trajectory E), but as seen in the example of Figure \ref{fig:hmf}, it mostly results in chaotic paths as seen in trajectories A and B. In this case, we believe these chaotic paths happen because when a pebble gets close to one planetesimal it stays in its vicinity for longer and has a high probability of getting slung around, which would result in a chaotic path. In other scenarios (different $\tau_\mathrm{s}, a_\mathrm{b}$, etc.), we can see pebbles spiralling in slowly and favouring hitting the outer (less massive) planetesimal, as in Figure \ref{fig:fidacc}.
    
In the second, as shown in the middle of the hitmap (in Figure \ref{fig:hmf}, specifically between $0.28R_\mathrm{H}\lesssim x_0\lesssim0.32R_\mathrm{H}$), the pebbles fall in from straight above the system. Falling in head on, with great velocity, either hitting one of the two bodies or going right past them and missing both. This region is exemplified by trajectories C and D.
    
In the third, shown on the right side, the pebbles come into the system circling around in retrograde fashion, opposing the movement of the binary. In the hitmap, this results in a large region of pebbles spiralling in slowly as seen in E. This happens all the way to $\sim 0.37R_\mathrm{H}$, from which the last trajectories are mostly chaotic paths. We believe this to result from the fact that the pebbles swirl around the binary and come back into the system almost from below. In that case, it would not have the same momentum as when it comes in from above since it now is moving opposite to the shearing wind. This lack of momentum makes it easy to get swung around when actually getting near the two bodies and do more chaotic paths, as exemplified in F.

\secondrev{Modifying the fundamental parameters $\tau_\mathrm{s}, a_\mathrm{b}$, and mass ratio will affect where exactly these three regimes (prograde, head on, and retrograde) are, but the fundamental structure remains similar.}

\begin{figure*}[t]
    \centering
    \includegraphics[width=1.\textwidth]{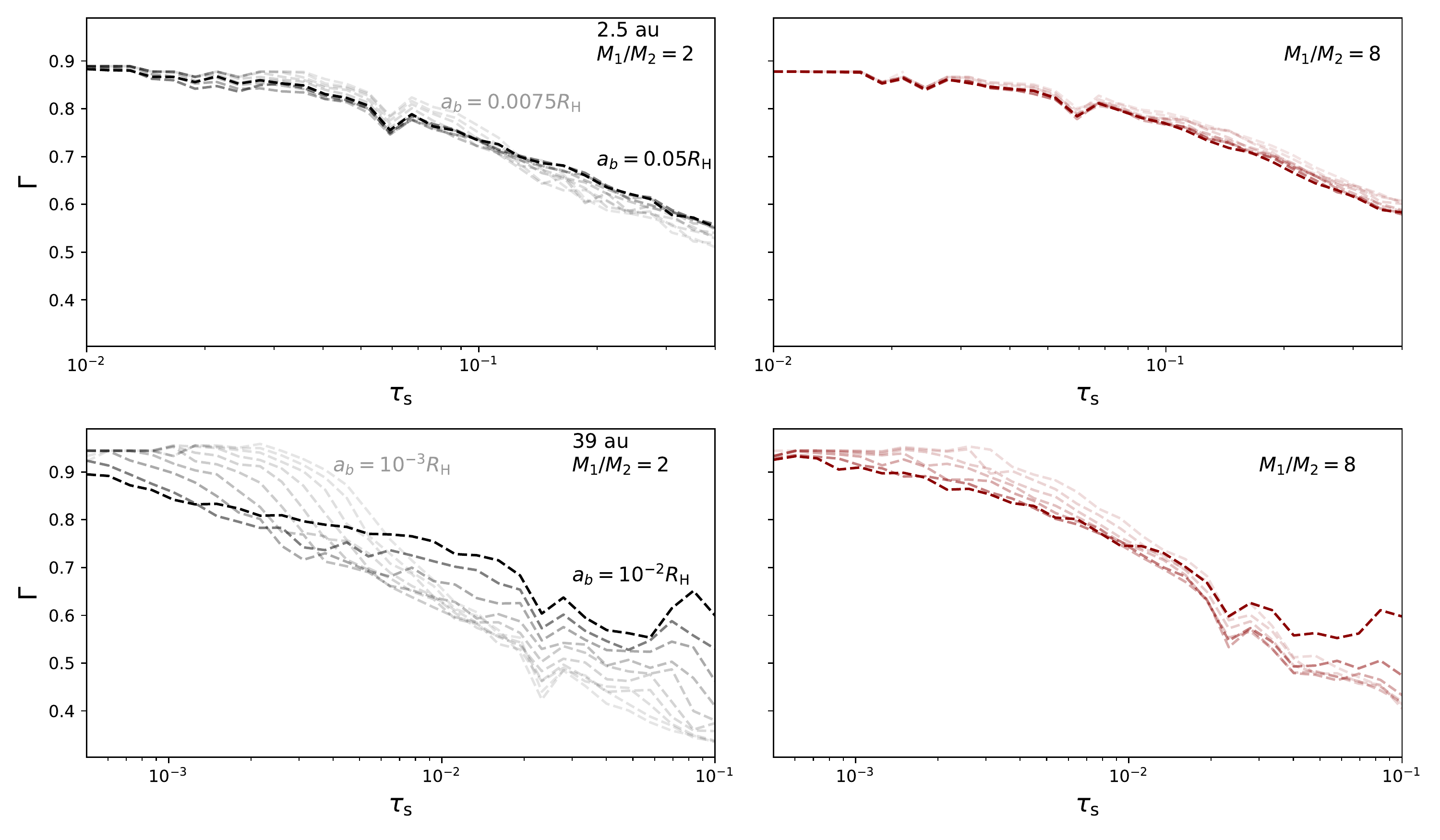}
    \caption{Growth efficiency, $\Gamma,$ as described in Equation \ref{eq:Gamma}. Left: Mass ratio of $f_m=2$. Right: Mass ratio of $f_m=8$. for a range of Stokes numbers, $\tau_\mathrm{s}$, mass-ratios, and binary separation, $a_\mathrm{b}$, at orbital distance of $r_0=2.5$ au from the central star (top panels) and $r_0 = 39$ au (bottom panels). The most opaque lines denote an orbital separation of $a_\mathrm{b}=0.0075R_\mathrm{H}$ while the most transparent lines shows an orbital separation of $a_\mathrm{b}=0.05R_\mathrm{H}$, other lines indicate orbital separations going logarithmic in between.}
    \label{fig:2-5aures1}
\end{figure*}

\subsection{Spiralling pebbles}
\label{sec:spiralling}
The situation where a pebble is caught by the binary into a spiralling motion is particularly interesting.
For an equal-mass binary, both bodies are then equally likely to accrete the pebble. However, if the binary has one body more massive than the other and if the distance between the two bodies is smaller than the initial size of the spiral, the lower-mass body becomes the more likely target. This important effect is illustrated in Figure \ref{fig:fidacc}. In this case, the masses are different, specifically, $M_1 = 2M_2$. Integrated over all pebbles in the impact parameter, the accretion ratio $\varepsilon$ as described in Equation \ref{eq:accratio} of this system is less than one, meaning that the smaller body is accreting more mass per unit time than its more massive counterpart. This may seem counter-intuitive as the less massive body has a smaller gravitational pull.  However, it can be understood by the in-spiralling motion of the pebbles that are bound to the binary system long before they actually get accreted. Figure \ref{fig:fidacc} shows a number of pebbles from a horizontal row in a hitmap. These pebbles have different starting positions $x_0$, but are launched with the same starting angle $\varphi_0$ of the binary. Many of the orbits turn into a spiral that slowly closes in on the binary. The small body moves like a vacuum cleaner through its orbit around the massive body (in fact, the barycenter, which will be very close to or even inside the larger body). As long as the settling velocity on the spiral path is low, the smaller body is likely to sweep up the incoming pebble before it can reach the massive body.

\begin{figure*}[t]
    \centering
    \includegraphics[width=1.\textwidth]{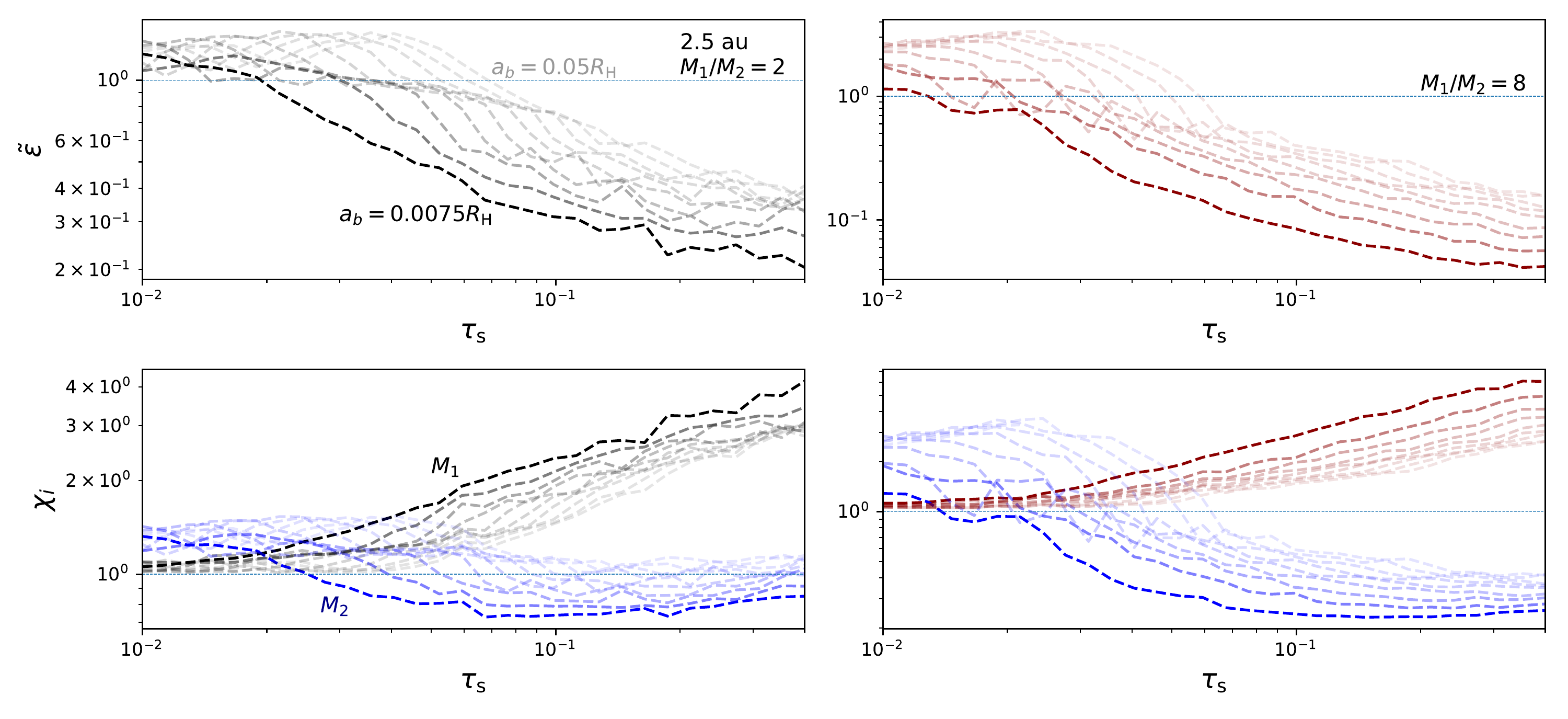}
    \caption{Normalised growth ratios expressed in $\Tilde{\varepsilon}$ (as described in Equation \ref{eq:epsilontilde}) for a range of Stokes numbers $\tau_\mathrm{s}$, mass-ratios, and binary separation, $a_\mathrm{b}$, at orbital distance of $r_0=2.5$ au from the central star, shown at the top. The most opaque lines denote an orbital separation of $a_\mathrm{b}=0.0075R_\mathrm{H}$, while the most transparent lines show an orbital separation of $a_\mathrm{b}=0.05R_\mathrm{H}$; other lines indicate orbital separations going logarithmically in between. Relative growth timescales of both bodies as a fraction of the growth timescale of a single body with mass, $M=M_1+M_2$ (as described in Eqs. \ref{eq:tgrowth1} and \ref{eq:tgrowth2}), shown at the bottom. Black and red lines denote $M_1$, while the blue lines denote $M_2$.}
    \label{fig:2-5aures}
\end{figure*}

\begin{figure*}[t]
    \centering
    \includegraphics[width=1.\textwidth]{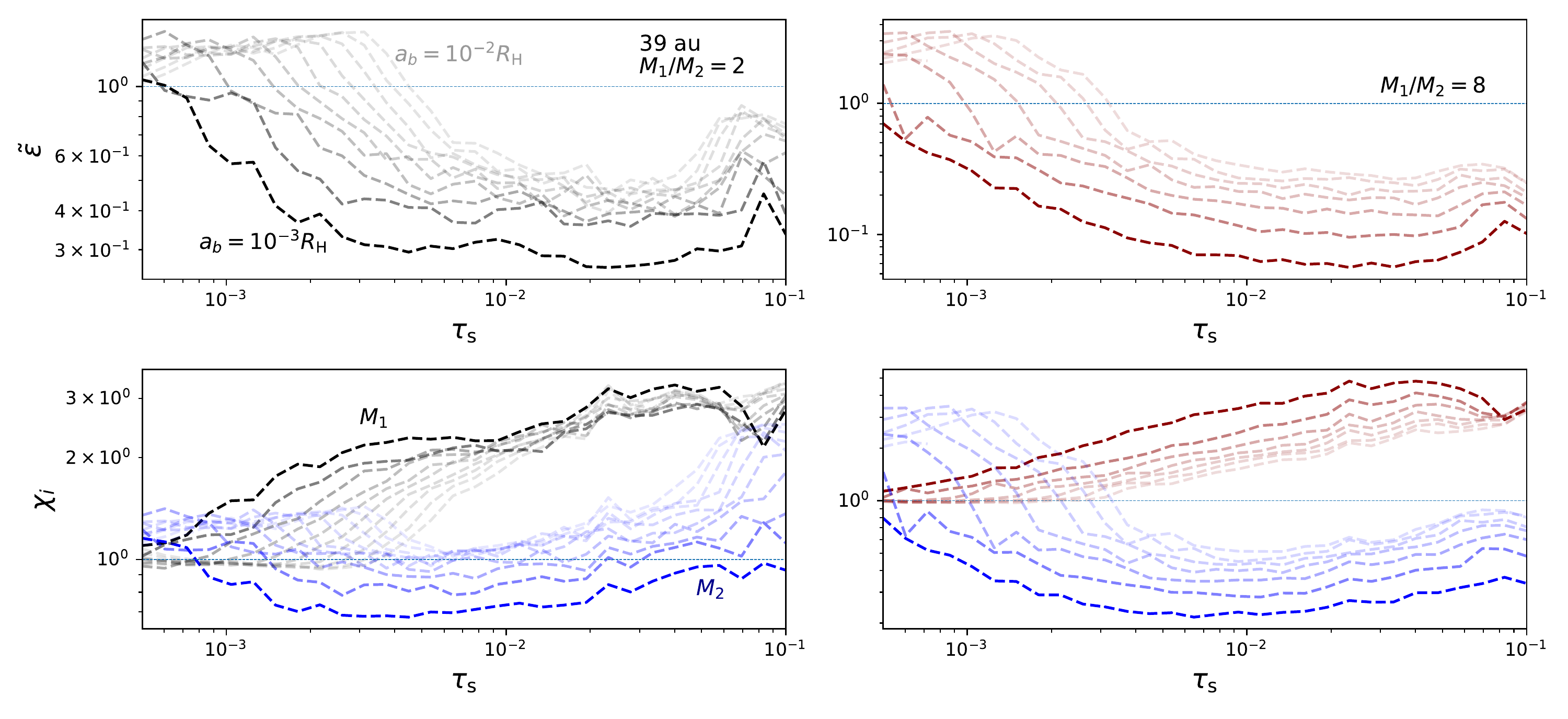}
    \caption{\secondrev{Normalised growth ratios expressed in $\Tilde{\varepsilon}$ (as described in Equation \ref{eq:epsilontilde}) for a range of Stokes numbers $\tau_\mathrm{s}$, mass-ratios, and binary separation, $a_\mathrm{b}$, at orbital distance of $r_0=39$ au from the central star, shown at the top. The most opaque lines denote an orbital separation of $a_\mathrm{b}=10^{-3}R_\mathrm{H}$, while the most transparent lines show an orbital separation of $a_\mathrm{b}=10^{-2}R_\mathrm{H}$; other lines indicate orbital separations going logarithmically in between. Relative growth timescales of both bodies as a fraction of the growth timescale of a single body with mass, $M=M_1+M_2$ (as described in Eqs. \ref{eq:tgrowth1} and \ref{eq:tgrowth2}), shown at the bottom. Black and red lines denote $M_1$, while the blue lines denote $M_2$.}}
    \label{fig:39aures}
\end{figure*}

\subsection{Accretion efficiency}
The impact parameter range of pebbles reaching the binary turns out to be the same as in a computation where the total mass of the binary is present as a single, larger body. This is the case when the system has a small binary separation compared to the range of impact parameters in which pebble accretion is relevant. Nevertheless, not all pebbles within the impact parameter range are actually accreted by the binary. The pebbles that are lost escape through scattering by close encounters with the primary and the secondary mass, as illustrated in Figure \ref{fig:hmf} (A, D, and F). This is different from the single mass case, in which, for the considered parameter space, all pebbles in the release range would be accreted. Therefore, we define the fraction of pebbles accreted on the binary, compared to the single-body case:
\begin{equation}
    \Gamma \equiv \frac{N_1 + N_2}{N_\mathrm{all}},
    \label{eq:Gamma}
\end{equation}
\secondrev{where $N_\mathrm{all}$ is the total number of released pebbles and $N_{1,2}$ are the number of pebbles accreted on bodies 1 and 2, respectively. For a single body with the total binary mass this fraction, $\Gamma$, would be unity.} The resulting $\Gamma$ values for our parameter study are shown in Figure \ref{fig:2-5aures1} for 2.5 au and 39 au. The accretion efficiency can drop to as low as $\Gamma \sim 0.4$ for the maximum Stokes numbers considered at 39 au. 

Knowing the accretion efficiency $\Gamma$ of the binary system, we can relate the growth time of the binary system to the growth time of a single body with the same mass. That is, in the pebble accretion regime, the growth timescale, $t_\mathrm{growth}$, of a single body with radius, $R$, internal density, $\rho_\bullet$, and mass, $M$, is given by \citep{Visser_2016}:
\begin{equation}
        t_\mathrm{growth} = \frac{M}{\dot{M}} = \frac{4\rho_\bullet R}{3v_\mathrm{hw}\rho_\mathrm{p}f_\mathrm{coll}},
        \label{eq:t_growth_single}
\end{equation}
where $\rho_\mathrm{p}$ is the spatial density of solid particles and $f_\mathrm{coll}$ \secondrev{the collision factor}. In the next section, we determine the relative growth timescales of the binary components based on this analysis.

\subsection{Relative growth rates}
To see if the binary mass ratio will shift towards unity during growth by pebble accretion, we need to look at the normalised growth ratio
\begin{equation}
    \label{eq:epsilontilde} \Tilde{\varepsilon} = \frac{M_2}{M_1}\varepsilon .
\end{equation}
Whenever this ratio $\Tilde{\varepsilon}$ dips below one, the binary will grow towards equal mass. In the top parts of Figures \ref{fig:2-5aures} and \ref{fig:39aures}, values of $\Tilde{\varepsilon}$ is shown as a function of the pebble Stokes number $\tau_\mathrm{s}$ for various combinations of our key parameters.
There is a clear regime, namely, an interval of Stokes numbers, where the mass ratio is indeed moving towards equal mass. This is both true in the asteroid belt at 2.5 au, and in the Kuiper belt at 39 au. This effect is seen not only for a moderate mass ratio of $f_m=2$, but also in the more extreme case of $f_m=8$.  It becomes more prevalent when the two objects in the binary are closer together. In Figures \ref{fig:2-5aures} and \ref{fig:39aures} it can be seen that while this is not always the case, it is true for a substantial part of the parameter study, in particular for large Stokes numbers. In this parameter regime, pebble accretion works toward reducing the mass ratio of the binary. The question arises as to how relevant this effect is. We consider whether bodies spend enough time in this regime to have their mass ratio significantly altered. In fact, it comes down to the question of what the growth timescales of these bodies are, \secondrev{where the growth timescale is the time to $e$-fold the mass of a component.} 

With the growth timescale derived for a single body and the accretion efficiency $\Gamma$ we can derive the mass accreted by each body over a set amount of time:
\begin{equation}
    \dot{M}_1 + \dot{M_2} = \Gamma\dot{M},
\end{equation}
which can be rewritten as:
\begin{align}
    \dot{M}_1 &= \frac{\Gamma\dot{M}}{\left(1+\varepsilon^{-1}\right)},\\
    \dot{M}_2 &= \frac{\Gamma\dot{M}}{\left(1+\varepsilon\right)},
\end{align}
since $\varepsilon = \dot{M}_1/\dot{M}_2$. We now can use the mass ratio, $f_m$, such that $M_1 = \frac{f_m}{1+f_m}M$ and $M_2 = \frac{1}{1+f_m}M$. In this way, we find the growth timescales of both planetesimals:
\begin{align}
    \label{eq:tgrowth1} t_\mathrm{growth,1} &= \frac{M_1}{\dot{M}_1} = \frac{\left(1 + \varepsilon^{-1}\right)f_m}{\left(1+f_m\right)\Gamma}t_\mathrm{growth} \equiv \chi_1t_\mathrm{growth},\\
    \label{eq:tgrowth2} t_\mathrm{growth,2} &= \frac{M_2}{\dot{M}_2} = \frac{\left(1 + \varepsilon\right)}{\left(1+f_m\right)\Gamma}t_\mathrm{growth} \equiv \chi_2t_\mathrm{growth},
\end{align}
where we have defined $\chi_i$ to be the scale factor that compares the $t_\mathrm{growth}$ of a single body to that of both planetesimals in a binary. The normalised accretion ratio $\Tilde{\varepsilon}$ and the growth timescales of both bodies are shown in the top and in the bottom panels of Figures \ref{fig:2-5aures} and \ref{fig:39aures}, respectively. At both 2.5 and 39 au (as we see in Figure \ref{fig:2-5aures1}),  the efficiency goes down with increasing, $\tau_\mathrm{s}$, which can also be seen with the accretion ratio $\varepsilon$. We believe both effects have somewhat the same cause: $\varepsilon$ favours the smaller body since it sweeps up the inspiralling pebbles from the larger mass (as described in Section \ref{sec:spiralling}), while the efficiency will also go down as the smaller mass throws a significant fraction of these inspiralling pebbles out of the system in close encounters.

\begin{figure*}[t!] 
    \centering
    \includegraphics[width=1\textwidth]{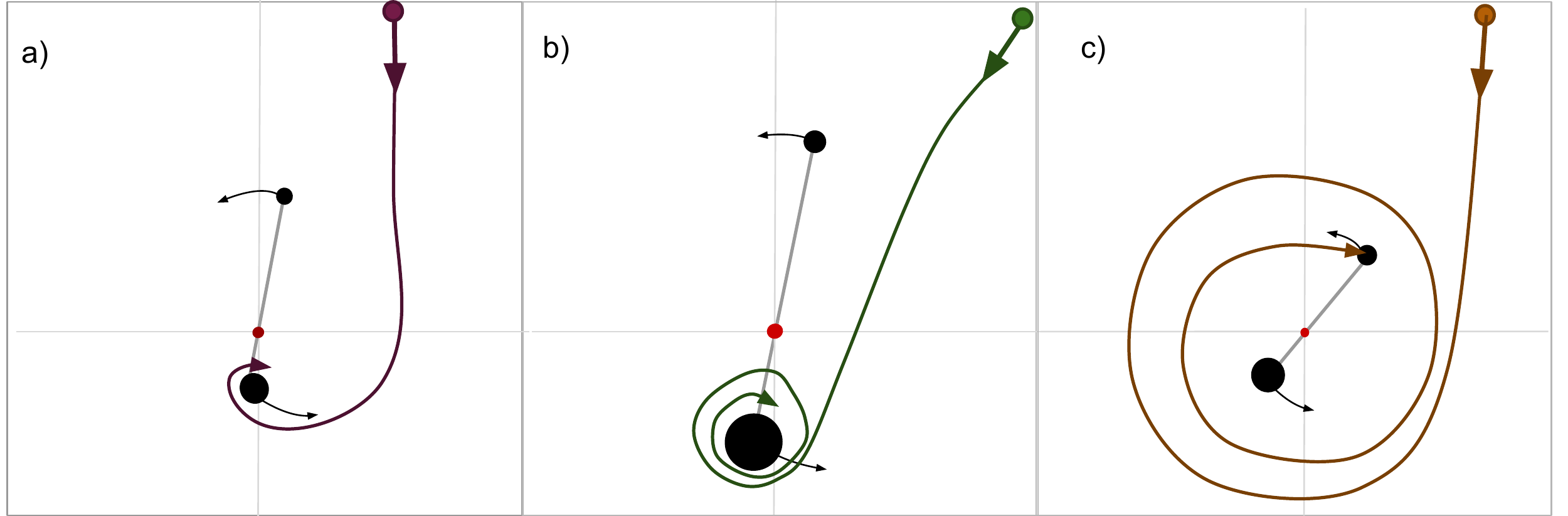} 
    \caption{Sketch of three distinctive cases that characterise the accretion efficiency of the binary components. The red dot indicates the binary barycenter, the small and large black dots the secondary and primary, respectively, and circular velocities shown by the black arrows (illustrative, and not shown to scale). Case a): Pebble Stokes number $\tau_\mathrm{s}$ is low and strongly coupled to the gas. It impacts the primary after typically a single encounter and renders the probability low that it will impact the secondary. Case b): Orbital separation, $a_\mathrm{b}$, is too wide for the secondary to benefit from accretion on the massive one. Case c): Narrow binary separation combined with the pebble accretion regime in which pebble spirals are ample and dominant. The probability of impact on the secondary is high due to the repeating encounters before the pebble sinks toward the massive body.}
    \label{fig:cases}
\end{figure*}

What can also be clearly seen is that the relative growth timescale of the larger body is greater than one in almost all of the parameter space, indicating that the growth of a massive body in the pebble accretion regime is slowed down when it has a secondary. In a large part of the parameter space, the $\chi_2$ value of the smaller body drops below 1, indicating that its growth will be accelerated by the presence of the larger partner. Even in the region where the efficiency $\Gamma$ decreases substantially, the growth of the smaller body is faster. That, combined with the fact the larger body grows more slowly, shows that we have a stabilising effect on the mass ratio. Similar-mass binaries are made more common at both 2.5 and 39 au by this process.

For Stokes numbers close to 0.1, which happens to be also the preferred Stokes number regime for triggering the streaming instability \citep{Baietal2010}, the growth timescales of the smaller body can easily be a factor of 3-5 less than that of the larger body.  So in the time the mass of the large body grows by a factor $e$, the mass of the smaller body may grow by a factor $e^3$ to $e^5$, and the mass ratio would change by a factor $e^2$ to e$^4$ or about 7 to 40.  This clearly demonstrates the strong potential of pebble accretion to create equal-mass binaries, even if the initial binary had demonstrated a high mass ratio.

\section{Discussion}
\label{sec:3}
\subsection{Pebble trajectory regimes}

We have identified three different main regimes that determine the enhancement in the accretion rate of the secondary (excluding chaotic paths that pebbles may take, as shown in Figure \ref{fig:hmf}A, B, and F):
\begin{enumerate}
    \item{\secondrev{The Stokes number, $\tau_\mathrm{s}$,} is too low and pebbles are tightly coupled to the gas. The pebbles do not spiral but make a direct impact or typically one loop before impact (Fig. \ref{fig:cases}a). The accretion ratio is in favour of the primary.}
    \item{When the binary separation is too wide, leading to the massive body accreting pebbles in isolation from the smaller one (Fig. \ref{fig:cases}b), the accretion ratio is in favour of the primary.}
    \item{When $\tau_\mathrm{s}$ is in the regime of pebble accretion and the separation falls within or on the typical radial extent of the spirals. The pebble undergoes tightly wound long-term spiralling inward and inevitably encounters the secondary, before reaching the more massive one (Fig. \ref{fig:cases}c). Here, the accretion ratio is in favour of the secondary.}
\end{enumerate}
Unsurprisingly, we find the same trend persists for varied orbital distance $r_0$. We show the final accretion efficiency for a distance of $r_0 = 2.5$ au and have verified it shows the same trend for $r_0 = 10$ au in the same manner as Figure \ref{fig:39aures}. While the parameters shift somewhat, the results show the same outcome.

\subsection{Timescale for developing toward equal mass}
\begin{figure}
    \centering
    \includegraphics[width=0.5\textwidth]{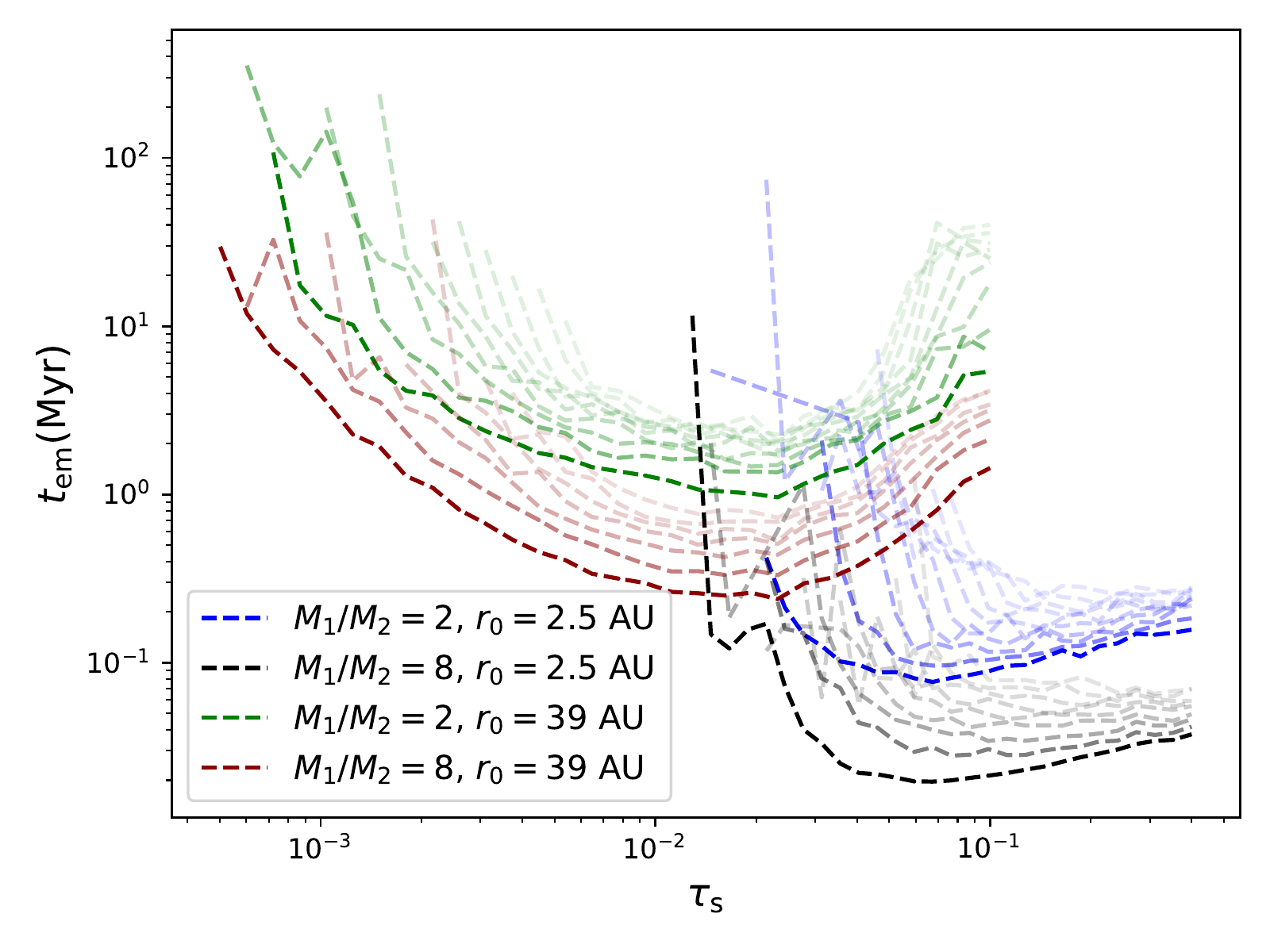}
    \caption{Timescale of $e$-folding towards equal mass $t_\mathrm{em}$ as described in Equation \ref{eq:t_em} for two tested mass ratios, $f_m=[2,8],$ at two different orbital distances, $r_0=[2.5,39]$ AU, and at different binary separations, $a_b$, denoted by the opaqueness of the lines. We note only the values, where $\Tilde{\varepsilon} < 1$ are plotted  since we are looking at the times it takes for the mass ratio, $f_m$, to grow towards unity, not away from it.}
    \label{fig:t_em}
\end{figure}

It is informative to see non-equal mass binaries growing towards a mass ratio of unity, but it does not give us a definitive timescale of how long it will take to get there. We consider now the timescale to change the mass ratio, $f_m$, by a factor $e$: 
\begin{align}
    \nonumber t_\mathrm{em} &= \frac{f_m}{\dot{f_m}} = \left(\frac{1}{1-\Tilde{\varepsilon}^{-1}}\right)\frac{M_1}{\dot{M_1}} = \frac{\Tilde{\varepsilon}}{\Tilde{\varepsilon}-1}t_\mathrm{growth,1}\\
    \label{eq:t_em} &= \frac{\Tilde{\varepsilon}f_m + 1}{\left(\Tilde{\varepsilon}-1\right)\left(1+f_m\right)}\frac{t_\mathrm{growth}}{\Gamma}.
\end{align}
with $\Tilde{\varepsilon} = \varepsilon / f_m < 1$ and both $\dot{f_m}$ and $t_\mathrm{em}$ being negative since we are looking at the time it takes to grow towards equal mass. This equal mass timescale $t_\mathrm{em}$ is plotted in Figure \ref{fig:t_em} for our parameter study. We are in an environment where the streaming instability is active or has been active recently, resulting in significant dust settling in the disk. We therefore take for the solid-to-gas ratio a conservative value of 0.1 such that $\rho_p$ from Equation \ref{eq:t_growth_single} is given by:
\begin{equation}
    \label{eq:solidtogasratio} \rho_p = 0.1\rho_g = 0.1 \frac{\Sigma}{\sqrt{2\pi}H}
,\end{equation}
where $\rho_g$ is the density of the gas (see Equation\ \ref{eq:rho}). We plot this $t_\mathrm{em}$ in Figure \ref{fig:t_em}. For a binary \myrev{with a total mass equal to that of the Pluto-Charon system} at the location of the asteroid belt at 2.5 AU, the timescales to $e$-fold towards unity are well within a Myr. Even at a greater distance of 39 AU, in a substantial part of our parameter space, we find timescales shorter than 1 Myr, again implying that pebble accretion has true potential as a formation method of (nearly) equal-mass binaries. Even though we have taken the static approach of calculating $t_\mathrm{em}$ by looking at a fixed, rather than evolving, mass ratio, from the figure we can conclude that $t_\mathrm{em}$ becomes lower with increasing mass ratio, $f_m$. Hence, the mass-equality timescale slows down while this process is happening, but initially the masses are rapidly converging.

\myrev{A question arises regarding whether the timescales in Figure \ref{fig:t_em} hold for smaller binaries than Pluto-Charon. Since $t_\mathrm{growth}$ of a single body at 3 AU is already within a few Myrs once its radius is larger than $\sim$300km \citep{Visser_2016}, we estimate the $t_\mathrm{em}$ of binaries with bodies between that and Pluto ($\sim$1200km) to be on the same order as shown in Figure \ref{fig:t_em}.}

\subsection{Limitations of the model}

One important simplification we have made is the assumption of 2D accretion, only looking at particles located in the midplane and ignoring the effects of a vertical distribution among incoming pebbles.  We can speculate that dropping this assumption may reduce the efficiency of the 'vacuum cleaning' by the quickly orbiting secondary.  An exploration of this effect would be very useful and should be done. For now, we hypothesise that even though the specific values of growth ratios may differ from the 3D results, the general trends will remain. The parameter space of Stokes numbers, $\tau_\mathrm{s}$, where the growth ratio flips in favour of the smaller body, will also remain largely the same. Therefore, we think the three main regimes of determining accretion rates we found in binaries will also remain the same as in our 2D case.

We are ignoring the effect of gas drag on the binary itself. While this assumption is certainly correct during the computation of the trajectory of a single pebble, over long evolutionary times it may affect the orbit of the binary by removing angular momentum from it.

In this work, we only considered purely circular orbits in our binary systems. Since truly circular orbits are rarely ever seen, our results might differ if we also simulate elliptical orbits. Most observed binaries that are in close orbits have been tidally circularised \citep{Nolletal2020} and we find that the assumption for circular orbits is justified. We estimate, however, that the trends will again stay largely the same for eccentric binaries due to the spiralling nature of pebbles in the 'die-hard' pebble accretion regime. 

We have only considered prograde orbital orientations for the binary mutual motion. The reason for this is that binaries forming in the context of the streaming instability show a clear preference for prograde orbits and is also the predominant preference in observations of the cold classical Kuiper belt population  \citep{Grundyetal2019,nesvorny2020binary,VisserandBrouwers2022}.

\section{Conclusions}
\label{sec:4}
In this paper, we study the growth and evolution of a binary planet (or planetesimal) system subject to pebble accretion. Specifically, we have investigated the accretion efficiency of the individual binary companions and the binary system as a whole for a variety of parameters such as Stokes number, binary separation, and binary mass ratio. 
If the binary components are pebble accreting with a mass ratio of unity, they accrete pebbles at an equal rate, as expected from symmetry arguments. If the mass ratio is not unity, we arrive at the following conclusions:

\begin{enumerate}
    \item A single planetary body that is accreting pebbles will generally accrete more efficiently in a binary system with a more massive companion, if pebbles are slowly spiralling toward the binary center-of-mass and these spirals extend to the length of the binary separation distance of $a_\mathrm{b}$ or larger.
    \item In that case, the mass ratio starts off above unity and the growth rate of the smaller body can be accelerated whilst the more massive one grows more slowly, pushing the binary system toward a mass ratio of unity. For Stokes numbers $0.01 \leq \tau_s \sim 0.1,$ the growth timescale of the smaller body can be three to five times shorter, changing the mass ratio quickly.
    \item Generally, three regimes can be distinguished: (i) Stokes numbers are too small and pebbles are tightly coupled to the gas. These pebbles impact the more massive body, typically after a single encounter, with no accretion advantage for the secondary. (ii)  The same holds when Stokes numbers are too large and pebbles meet the binary on ballistic orbits. (iii) Pebbles are captured by the primary in orbits wider than the binary separation. During the in-spiralling, the pebble has a high probability to be swept up by the secondary, resulting in a converging mass ratio.
\end{enumerate}

\begin{acknowledgements}
We would like to thank J. Groot, G. Koning, R. M. van der Linden \& M. Rustenburg for their insightful comments and discussions. We thank everyone at the disk meeting for feedback and improvements to the manuscript. RGV acknowledges funding from the Dutch Research Council, ALWGO/15-01.
\end{acknowledgements}

% WARNING
%-------------------------------------------------------------------
% Please note that we have included the references to the file aa.dem in
% order to compile it, but we ask you to:
%
% - use BibTeX with the regular commands:
%   \bibliographystyle{aa} % style aa.bst
%   \bibliography{Yourfile} % your references Yourfile.bib
%
% - join the .bib files when you upload your source files
%-------------------------------------------------------------------
\bibliographystyle{aa}
\bibliography{references}
\end{document}